\documentclass[epj]{webofc}
\usepackage[varg]{txfonts}   
\wocname{EPJ Web of Conferences}
\woctitle{ICNFP 2016}
\newcommand{\numunue}{\ensuremath{\nu_\mu \rightarrow \nu_e}\xspace}
\newcommand{\nue}{\ensuremath{\nu_{e}}\xspace}
\begin{document}
\selectlanguage{english}
\title{The next challenge for neutrinos: the mass ordering}
%
%

\author{Luca Stanco\inst{1}\fnsep\thanks{\email{luca.stanco@pd.infn.it}} 
}

\institute{I.N.F.N. Padova, Via Marzolo, 8 I-35131 Padova, Italy
}

\abstract{%
Neutrino physics is nowadays receiving more and more attention as a possible source of information for the long--standing
investigation
of new physics beyond the Standard Model. This is also supported by the recent change of perspectives in neutrino
researches since the discovery period is almost over and we are entering
the phase of precise measurements. Despite the limited statistics collected for some variables, the three--flavour neutrino framework 
seems well strengthening. However some relevant pieces of this framework are still missing. The amount of
a possible CP violation phase and the mass ordering are among the most challenging and probably those that
will be known in the near future. In this paper we will discuss these two correlated issues and a very recent
new statistical method introduced to get reliable results on the mass ordering.

}
\maketitle
\section{Introduction}
\label{intro}
The current scenario of the Standard Model (SM) of particle physics, being arguably stalled by the discovery of the Higgs boson,
is {\em desperately} looking for new experimental inputs to provide a more comfortable theory.
In parallel, experiments on neutrinos so far have been  an outstanding source of novelty and unprecedented 
results. In the last two decades several results were obtained 
studying atmospheric, solar or reactor neutrinos, and more recently neutrino productions 
from accelerator--based beams. Almost all these results have contributed to strengthen 
the flavour--SM. 
Nevertheless, 
relevant parts like the leptonic CP phase and the neutrino masses are still missing, a critical ingredient
being the still undetermined neutrino mass ordering.
On top of that the possibility of lepton flavour violation (e.g. if neutrinos are Majorana particles) and the absolute masses 
of neutrinos are very open issues. Unfortunately these latter questions will probably take sometime to be answered, 
while the previous ones are matter of debates and experimental proposals. 

The achievements of the last two decades brought out a coherent picture,
namely the mixing of three neutrino flavour--states, $\nu_e$, $\nu_{\mu}$ and $\nu_{\tau}$, with three  $\nu_1$, $\nu_2$ and $\nu_3$ 
mass eigenstates. 
In the three--neutrino framework still an unknown parameter, highly correlated with the masses and
the CP angle, $\delta_{CP}$, is strongly pursed by current and proposed experiment, i.e. the neutrino mass ordering of the neutrino
mass eigenstates. Specifically, it is still largely unconstrained the sign of $|\Delta m_{31}^2|=|m_3^2 - m_1^2|$, the  difference 
of the squared masses of $\nu_3$ and $\nu_1$. The mass ordering (MO) is usually identified either as normal hierarchy (NH) 
when $\Delta m_{31}^2 > 0$ and inverted 
hierarchy (IH) in the opposite case. Its importance is enormous to provide inputs for the next studies and
experimental proposals, to finally clarify the needs or not for new projects, and to constraint analyses in other fields
like cosmology and astrophysics.

\begin{figure}[h]
\centering
\vspace{-1cm}
\includegraphics[width=8cm,angle=-90,clip]{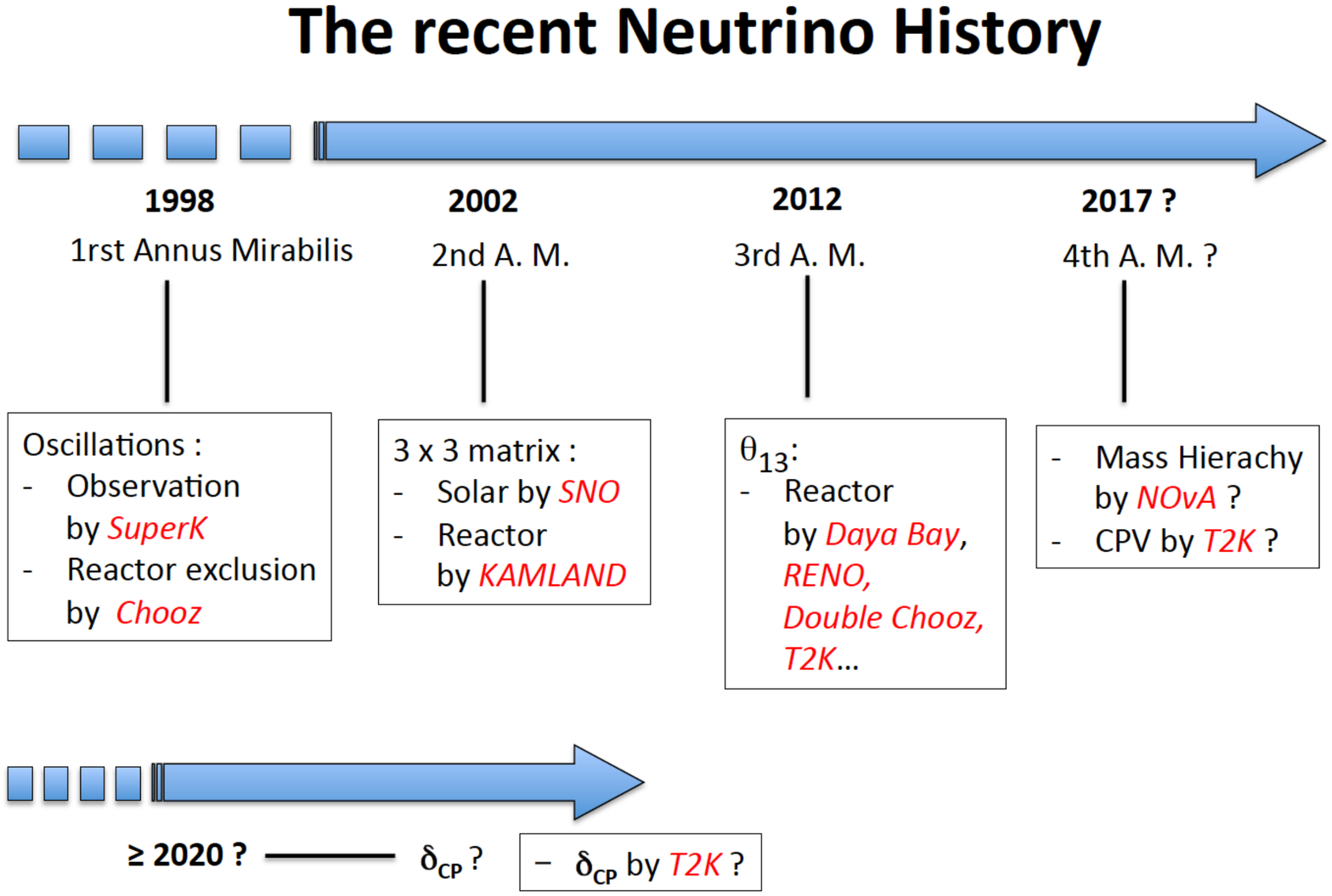}
\vspace{-1cm}
\caption{A synthetic view of the recent history of oscillation neutrino discoveries. Prospects for the next future are also drawn.}
\label{fig-1}       
\end{figure}

From the recent review~\cite{lucas} a synthetic flow chart can be extracted (figure~\ref{fig-1}), where some key steps have been
identified, the {\em anni mirabiles}. From the picture three specific years were fundamental to understand the neutrino paradigm:
1998, when the oscillations were first discovered by Super-Kamiokande~\cite{superk}, not forgetting the missing-observation of 
Chooz~\cite{chooz}; 2002, when the discovery of the solar oscillations by SNO~\cite{sno}, together with the oscillation pattern of 
reactors' antineutrino  measured by KAMLAND~\cite{kamland}, opened the way to the $3\times 3$ mixing scenario; 2012, 
when $\theta_{13}$, the third mixing angle, was finally measured
by Daya-Bay, Reno and Double-Chooz reactor experiments~\cite{dayabay} and T2K~\cite{T2K-th13}, an accelerator beam experiment. 
Next year might be the fourth key--year since NOvA provided first indications~\cite{nova-2015} on the mass ordering, and 
T2K recently released~\cite{T2K-prel} exclusion of $\delta_{CP}= 0$ at 90\% C.L..
The issue of the measurement on the mass ordering will be further discussed in this paper, 
while a robust measurement of $\delta_{CP}\neq 0$ could be soon obtained. Finally, from the year 2020 T2K
should begin a campaign of measurements on $\delta_{CP}$~\cite{T2K-future}.

\section{Neutrino mass ordering}
\label{mass-order}
\begin{figure}[h]
\centering
\vspace{-4cm}
\includegraphics[width=8cm,clip]{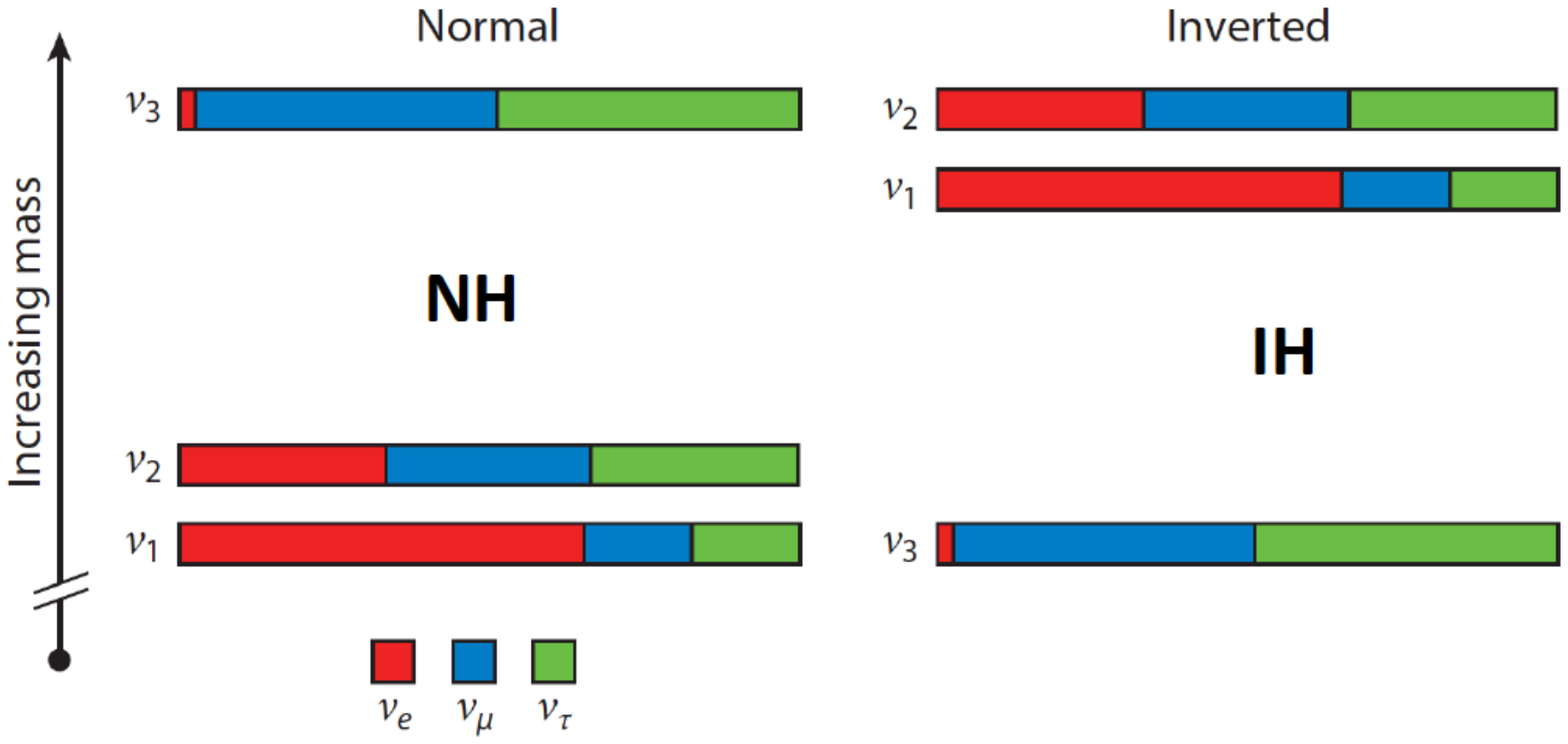}
\vspace{-4cm}
\caption{Illustration of the neutrino mass ordering for the two hypotheses, Normal Hierarchy (NH) and Inverted Hierarchy (IH). The 
flavour sharing due to the mixing is also drawn for each mass eigenstate.}
\label{fig-2}       
\end{figure}

The neutrino mass ordering indicates whether there are one or two highest standard-neutrino masses, as shown in figure~\ref{fig-2}
that illustrates the relative flavour sharing due to their mixing.
Sensitivities to the mass ordering are given by the contemporary presence of two or more contributions in the neutrino propagator (see 
e.g.~\cite{propag}). 
Specifically, as far as oscillations are concerned, neutrino mass hierarchy can be measured observing the interference of oscillations 
driven by $\pm \Delta m_{31}^2$ with oscillations driven by another quantity $Q$ with known sign. For example the dependence on MO 
at the long-baseline experiments with neutrino beams rises up from the interference terms between the pure flavour oscillation
\numunue, and the $Q$ 
term that describes the resonant flavour transitions when neutrinos propagate in a medium with varying density. The high statistic 
atmospheric experiments like INO (India)~\cite{india-ino}, KM3/ORCA (Europe)~\cite{km3} and PINGU (USA)~\cite{pingu} 
can access the MO based on the same feature. 
The same occurs for the collective effects in neutrino coming from supernovae explosions. Instead, the JUNO (China) experiment~\cite{juno}
foresees to establish the MO by looking at the interference between the solar and the atmospheric oscillations, i.e. the oscillations driven 
by $\Delta m_{21}^2=m_2^2 - m_1^2$ and $\Delta m_{31}^2$, respectively.	

\begin{figure}[h]
\centering
\includegraphics[width=8cm,clip]{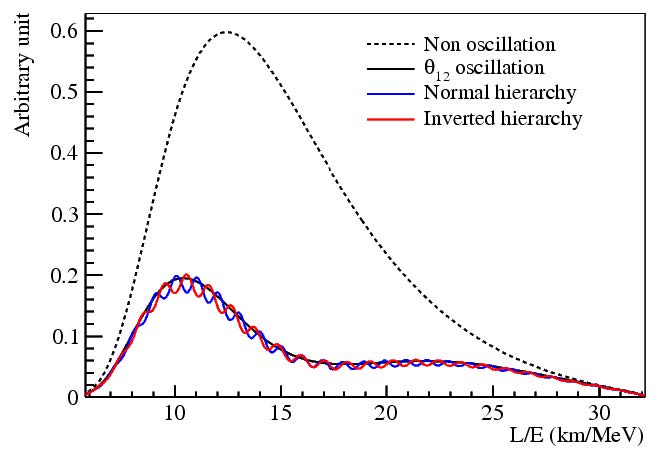}
\vspace{-0.5cm}
\caption{The relative shape difference of the reactor antineutrino flux for the two different neutrino mass ordering as function
of L/E (baseline distance L over the neutrino energy E). The large discrepancy with the non-oscillation expectation is due to the
solar oscillation, while the ripples are due to the atmospheric one. Taken from~\cite{juno-phys}.}
\label{fig-3}       
\end{figure}

\begin{figure}[hbt]
\centering
\includegraphics[width=11cm,clip]{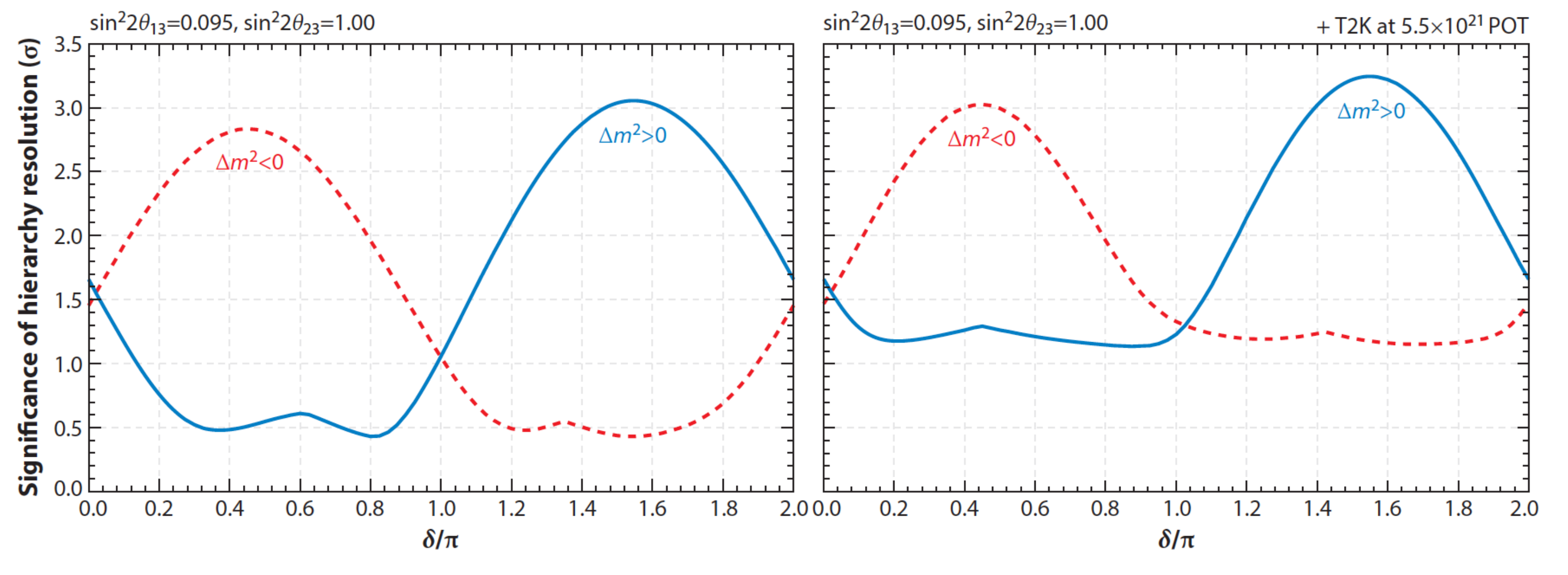}
\caption{Prospect of the mass ordering disentanglement after several years (3 with neutrino and 3 with antineutrino) 
of NOvA data taking, as function of $\delta_{CP}$ (left).  On the right the slight improvement in significance obtained by adding
few years of T2K exposure is shown. Taken from~\cite{nova-patt}.}
\label{fig-4}       
\end{figure}

For example in figure~\ref{fig-3},  the interference between the solar and atmospheric oscillations
brings up the ripple behaviour and its relative shift for the two options, NH and IH. The separation between the two patterns is 
manly matter of energy resolution. JUNO aims to achieve less than a 3\%/$\sqrt{E(MeV)}$ uncertainty. 

For the accelerator basis searches the NOvA
experiment foresees some information be available after several years of running (figure~\ref{fig-4}). 
Adding measurements on $\delta_{CP}$ from few years of T2K exposure 
will allow to slightly increase the separation between the two options in different portions of $\delta_{CP}$ range
(a nice review on the combined challenging for the current running experiments is given in~\cite{nagaya}).
However the expectation is not exciting, only a 3-sigma significance could be obtained and only in some favorable
$\delta_{CP}$ regions.

\begin{figure}[hbt]
\centering
\includegraphics[width=11cm,clip]{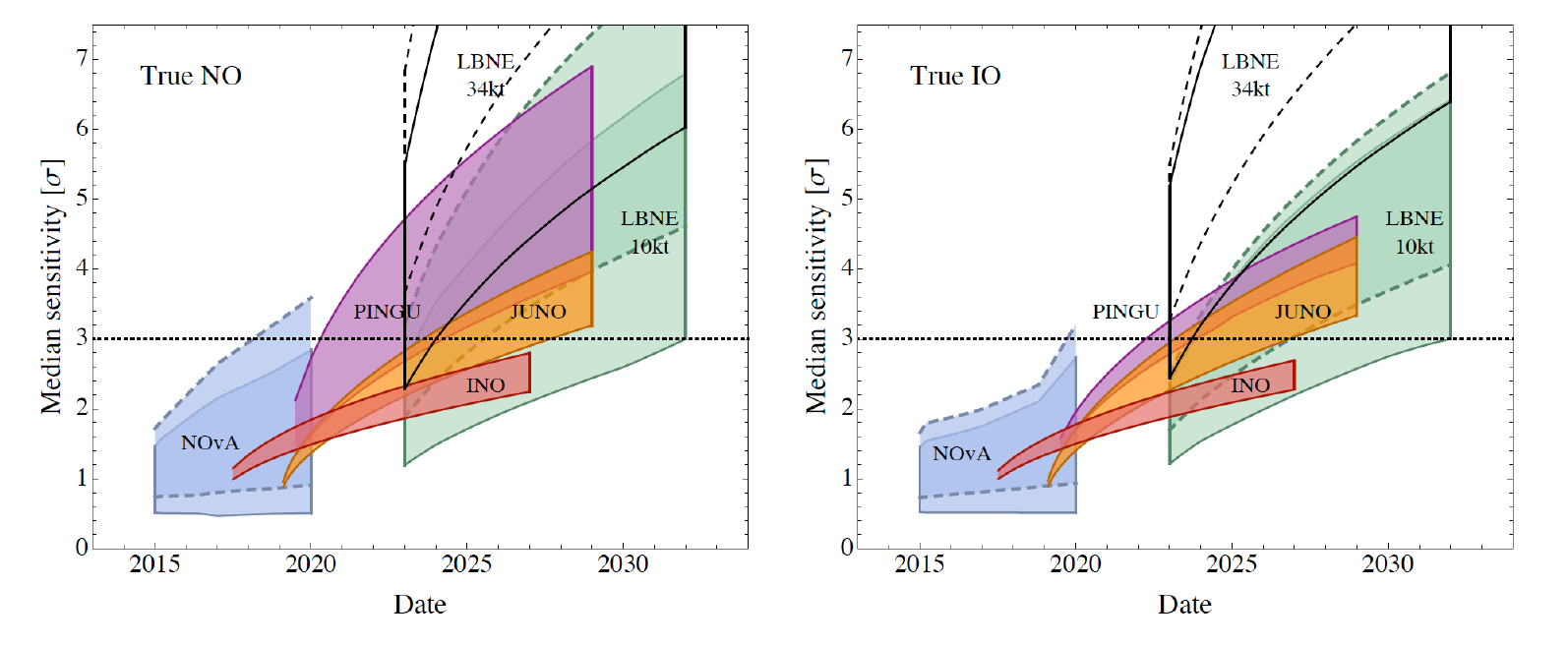}
\caption{Perspectives for the neutrino mass ordering in the next future.
The left (right) panel shows the median sensitivity in number of sigmas for rejecting the IO
(NO) if the NO (IO) is true for different facilities as a function of the date. The width of the bands
correspond to different  true values of the CP phase $\delta$ for NOvA and LBNE (now DUNE~\cite{dune}),
different true values of $\theta_{23}$ between 40$^0$ and 50$^0$ for INO and PINGU, and energy resolution between
3\%/$\sqrt{E(MeV)}$ and 3.5\% for JUNO. For the long baseline experiments, the bands with solid (dashed)
contours correspond to a true value for $\theta_{23}$ of 40$^0$ (50$^0$). 
NO (IO) stands for NH (IH). Taken from~\cite{blennow}.}
\label{fig-5}
\end{figure}

As a matter of fact the perspectives for the determination of the neutrino mass ordering in the near future are rather 
poor, even less favorable than the prospects for the $\delta_{CP}$ measurement. The overall scenario is summarized in 
figure~\ref{fig-5}\footnote{The recent update given by A. Heijboer~\cite{now2016} at the NOW2016 conference does not 
really change the perspectives, but a shortening  of the period needed to PINGU/ORCA to reach a high sensitivity:
starting hopefully in 2020 only 4 years instead of 8 would be needed to get around 5 sigmas.
See also the recent ARCA/ORCA Letter of Intent~\cite{km3}.}. 

There are concerns for each of the facilities:
\begin{itemize}
\item NOvA would need some luckiness;
\item PINGU (and ARCA/ORCA) are not yet fully funded;
\item INO would not reach a good significance;
\item JUNO would cope with a very challenging energy resolution;
\item DUNE should be able to get it in few years but it is not clear if and when it would start.
\end{itemize}

Thus it becomes highly worth to look for new techniques on analyses of the mass ordering in order to improve the
significance level and to gain in time schedule.
Following the discussion in~\cite{lucas} a change of perspective is first of all needed. 
Technically one should focus on the rejection of the wrong hierarchy and not on the observation of the
true one. Therefore it is mandatory to introduce new test statistics that would allow to distinguish between
NH and IH within that approach. 

Moreover it is also time to work out a comprehensive handling of all 
the future measurements on MO.
Reasons to alternatively obtain a robust result from only one experiment
are due to the lack of confidence on the 3-neutrino framework and/or the cross-correlation of the systematic errors 
between different experiments. The first concern should be targeted with specific experiments and
it should not attain the extraction of the oscillation parameters. The second concern about the systematic errors should
not prevent to conservatively use one experiment as a guideline adding information from the other ones in a
controlled way.

In the second part of this paper focus will be given on a very recent new statistical technique~\cite{lucas-mh}
that could be extensively applied to single or multiple measurements of the neutrino mass ordering.

\section{Standard techniques for MO}
\label {mh-stand}
Provided the current knowledge of the oscillation parameters~\cite{lisi2016}, with
uncertainties from few percents to more than 10\%, 
all the methods developed so far for 
establishing whether MO is normal or inverted are based on the computation of the $\chi^2$ difference with respect to the 
best fit solutions for the NH and IH  cases~\cite{mh-all}. These analyses make use of the test statistic

{\footnotesize
\begin{equation}
\Delta\chi^2_{min}= \chi^2_{min}(IH)-\chi^2_{min}(NH),
\end{equation}}
\noindent where the two minima are evaluated spanning the uncertainties of the three-neutrino oscillation parameters,
namely $\Delta m_{21}^2$, $\pm\Delta m_{31}^2$, $\theta_{12}$, $\theta_{23}$, $\theta_{13}$ and $\delta_{CP}$,
$\theta_{ij}$ ($i,j=1,2,3$) being the mixing angles in the standard parameterization.
The statistical significance in terms of $\sigma$'s is  computed as $\sqrt{\Delta\chi^2}$. The limits of such procedures
are well known~\cite{ciufoli}. In particular, the significance corresponds only to the median expectation and does not consider the
intrinsic statistical fluctuations. Thus errors of type I and II~\cite{pdg} should be taken into account when comparing the probability density functions
of each $\chi^2_{min}$, the corrected significance is lower and more sigma's have to be gained to reach a robust observation. 
Despite this and other caveats no alternative test statistic has been  outlined so far.

\begin{figure}[hbt]
\centering
\includegraphics[width=8cm,clip]{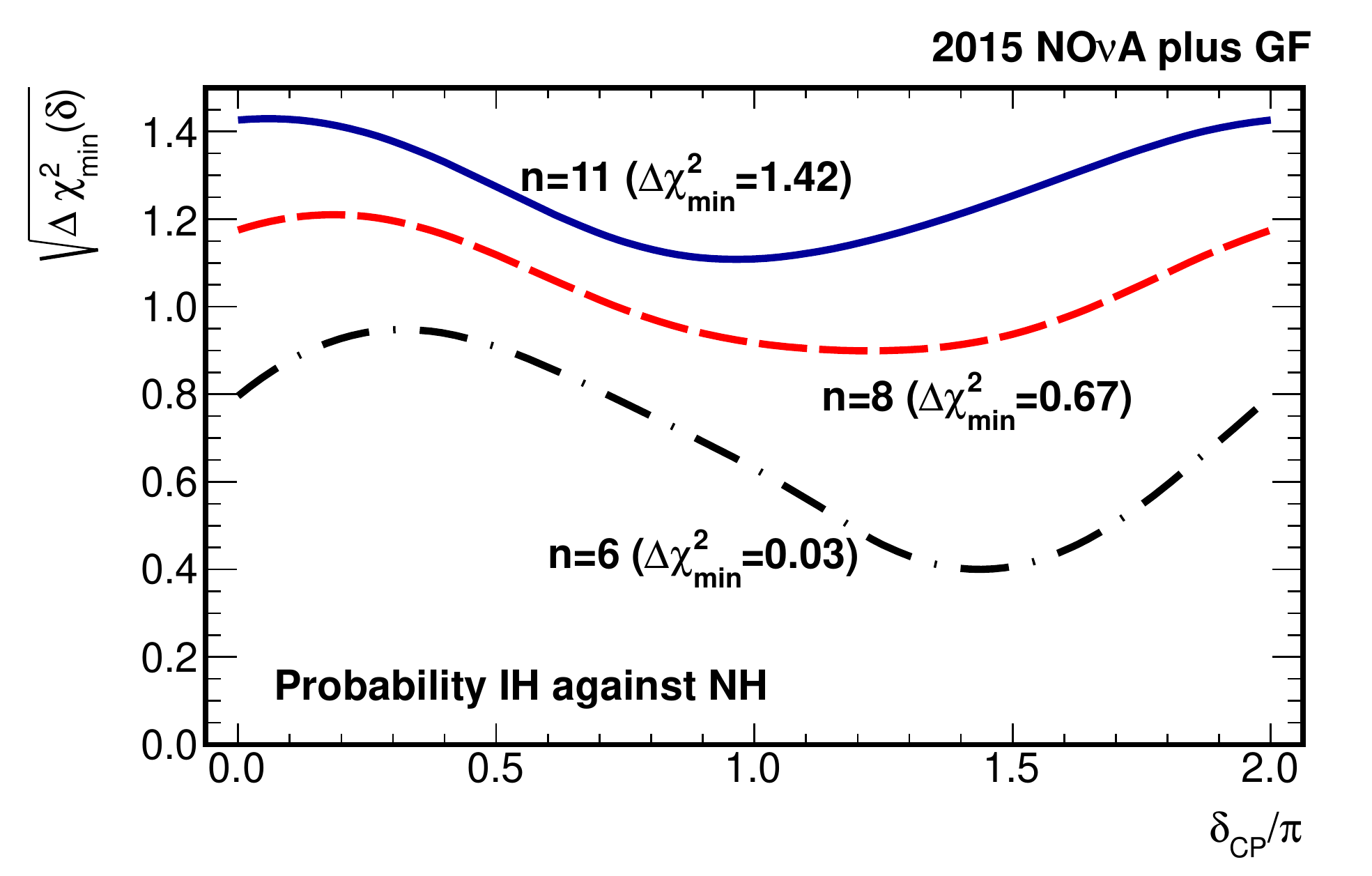}
\caption{The standard technique based on $\Delta\chi^2_{min}$ applied to the 2015 NOvA results. 
The significance of IH exclusion as function of $\delta_{CP}$ is shown.
The three curves correspond to the significances
for different numbers of observed events (6, 8 or 11) by NOvA in the \numunue channel, for a total exposure of
$2.74\times 10^{20}$ p.o.t.
Minimization over $\theta_{23}$ and $\theta_{13}$ and use of the best fit values from the global fit (GF)
analysis~\cite{lisi2016} for the other parameters, are performed. The quoted $\Delta\chi^2_{min}$ correspond to 
$\chi^2_{min}(IH)-\chi^2_{min}(NH)$ computed in the whole range of $\delta_{CP}$. Taken from~\cite{lucas-mh}.
}
\label{fig-6}
\end{figure}

Meanwhile first results from NOvA became available last year even if only with an exposure of
$2.74\times 10^{20}$ p.o.t.~\cite{nova-nue}. The standard analysis via $\Delta\chi^2_{min}$ on mass ordering performed on the 2015 NOvA 
\numunue appearance is reported in figure~\ref{fig-6}. It is confirmed that no real sensitivity is available yet on MO.

\section{A new techniques for MO}
\label {mh-new}

\begin{figure}[htb]
\centering
\includegraphics[width=0.8\linewidth]{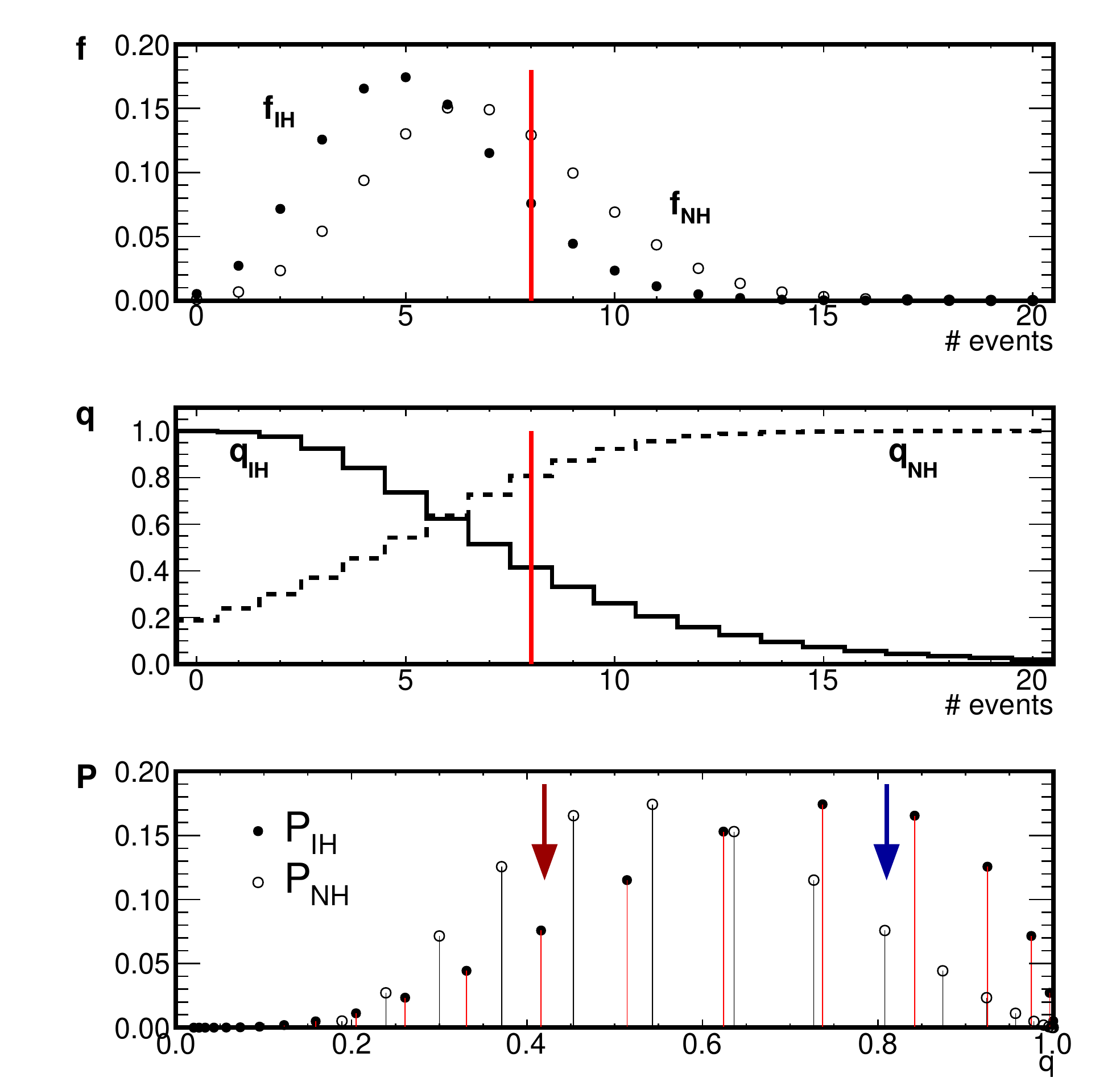}%
\caption{\label{fig-7} The behaviours of the distributions related to the new estimator $q_{MO}$.
Top: the expected Poisson distributions (signal plus background) for IH (plain dots) and NH (void dots), 
for $\delta_{CP}=3/2 \,\pi$ and $2.74\times 10^{20}$ p.o.t., in the 2015 NOvA analysis.
Middle: the corresponding values assumed by
$q_{IH}$ (plain line) and $q_{NH}$ (dashed line). Bottom: the probability mass functions of $q_{IH}$ (plain dots) and $q_{NH}$ (void dots).
The arrows indicate the thresholds used to compute $q_{IH}$ and $q_{NH}$ for $n=8$. Taken from~\cite{lucas-mh}. }
\end{figure}

The new technique reported in~\cite{lucas-mh} is based on a new test statistic that properly weights the intrinsic statistical 
fluctuations of the data and extracts the significances of either NH or IH.
First, the Poisson distributions for $n_i$ observed events $f_{MO}(n_i;\mu_{MO} |\delta_{CP})$ are considered,
where $\mu_{MO} (\delta_{CP} )$ are the expected number of events as function of $\delta_{CP}$,
 $MO$ standing for $IH$ or $NH$. 
For a specific $n$ the left and right cumulative functions of $f_{IH}$ and $f_{NH}$ are computed and their ratios, $q_{MO}$, are evaluated. 
The ratios are similar to the CL$_s$ test statistic~\cite{cls} used for the Higgs discovery. Since for the \nue appearance at NOvA the 
number of expected events as function of $\delta_{CP}$ is asymmetric towards IH and NH (less events are expected for IH than for NH),
the CL$_s$ are defined for either the IH or the NH case:

\vspace{-0.1cm}
 {\footnotesize
\[
q_{IH}(n, \delta_{CP}) =\frac{\sum_{n_i^{IH}\ge n} f_{IH}(n_i^{IH};\mu_{IH} |\delta_{CP} )}{\sum_{n_i^{NH}\ge n}
f_{NH}(n_i^{NH};\mu_{NH} |\delta_{CP} )},\quad
q_{NH}(n, \delta_{CP}) = \frac{\sum_{n_i^{NH}\le n} f_{NH}(n_i^{NH};\mu_{NH} |\delta_{CP})}{\sum_{n_i^{IH}\le n} 
f_{IH}(n_i^{IH};\mu_{IH} |\delta_{CP})}. 
\]
}
$q_{IH}$ and $q_{NH}$ are two discretized random  variables comprised to the [0, 1] interval. As $n$ goes to zero  $q_{IH}$
goes to one, while when $n$ increases $q_{IH}$ asymptotically tends to zero. $q_{NH}$ behaves the other way around 
towards $n$. For illustration purpose the behaviours of $f_{MO}$ and $q_{MO}$ are shown in Fig.~\ref{fig-7} for a typical case
($n=8$).

The probability mass functions, $P(n)$, of each $q_{MO}$
have been computed via toy 
Monte Carlo simulations based either on $f_{IH}$ (test of IH against NH) 
or $f_{NH}$ (test of NH against IH).
They are further compared to the observed data $n_D$, the number of observed events either in real data or in Monte Carlo simulation,
evaluating the $p$--value probabilities for $n_D$:\footnote{It is worthwhile to note that the method does not really need the double definition of $q_{IH}$ and  $q_{NH}$
since they are complementary. All the results can be obtained using only one expression, taking
the $p$--value in the proper domain. The procedure described in the text is at ease of the
reader.}

 {\footnotesize
\[
p_{IH}(n_D, \delta_{CP}) = \sum_{q'_{IH}\le q_{IH}(n_D)} P_{IH}(q'_{IH} |\delta_{CP}),\quad\quad
p_{NH}(n_D, \delta_{CP}) = \sum_{q'_{NH}\le q_{NH}(n_D)} P_{NH}(q'_{NH} |\delta_{CP}).
\]}
Finally the significance has been computed with the one--sided option since
this corresponds to 0 sigma ($Z=0$) for $p=50\%$ that equalizes the IH and NH probabilities.
Within that choice $Z$ is defined as $Z=\Phi^{-1}(1-p)$, where $\Phi^{-1}$ is the quantile of the standard Gaussian and Z is the number
of standard deviations. 

With the new method an averaged increase  of 0.5 $\sigma$ with respect to the standard $\Delta\chi^2_{min}$ is obtained~\cite{lucas-mh}.
Worth to note that the increase is not constant but it depends on the discrimination threshold $n_D$ and $\delta_{CP}$: 
the gain of the new method in terms of the number of sigma's strongly raises with $n_D$ and ``favorable''
regions of $\delta_{CP}$. As demonstrated in the appendix of~\cite{lucas-mhv2} the new method is generally better than $\Delta\chi^2_{min}$
for many reasons: it deals with the full probability distributions, it profits of the intrinsic fluctuations of the data and, most relevant,
it looks to the right question (to disprove one MO option). In fact the new $q$ estimator focusses on the possibility to reject
the wrong hierarchy, disregarding the other one. Therefore, once one option is selected (e.g. rejection of IH) it does not provide any 
evaluation on the other option (rejection of NH). Instead the $\Delta\chi^2_{min}$ method  treats the two option
in a symmetric way with the disadvantage of mixing up the information.

\begin{figure}[hbt]
\centering
\includegraphics[width=8cm,clip]{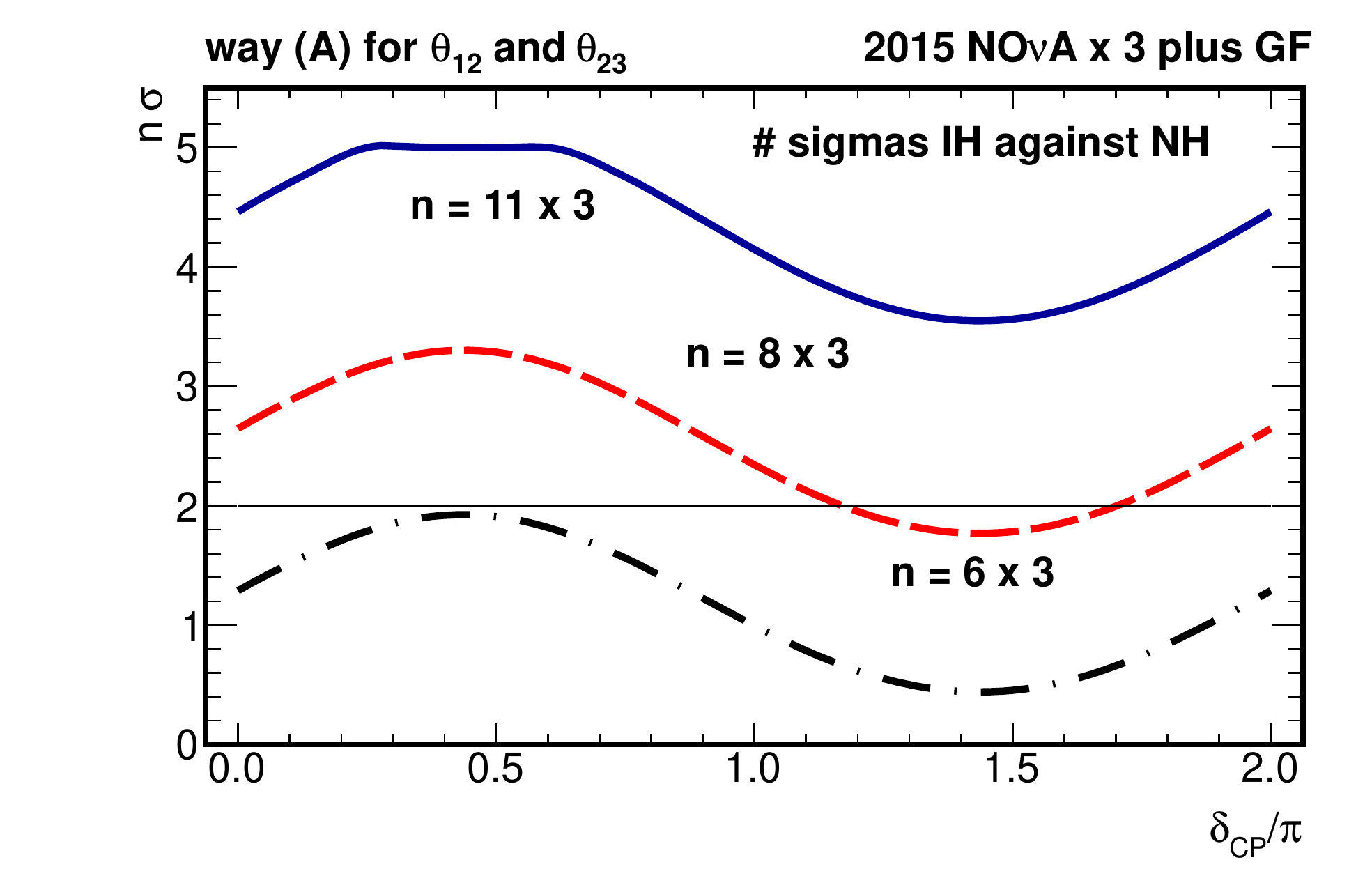}
\caption{The significance from the the new method in terms of $\sigma$ (one-sided option) is shown for a total exposure of
$8.22\times 10^{20}$ p.o.t. (top), assuming the same efficiency for the signal and the same level of background rejection
of 2015 NOvA analysis. The tested hypothesis is IH against NH. 
The different curves correspond to different possible observations of events, 18, 24 and 33 (signal plus background), as 
estimated by 2015 NOvA in the \numunue channel. The uncertainties on  $\theta_{23}$ and $\theta_{13}$ are treated as nuisances
(way A). Taken from~\cite{lucas-mh}.
}
\label{fig-8}
\end{figure}

Results are much more promising when the data sample increases. For example if a factor 3 in
exposure is applied to the 2015 NOvA analysis  the rejection of IH can reach the 95\% C.L. in almost the full range of $\delta_{CP}$, even
including the current uncertainties on $\theta_{23}$ and $\theta_{13}$ in a Bayesian way (figure~\ref{fig-8}).
The new method works better when some statistical fluctuations with respect to the median expectations
should be observed.
The 95\% C.L. exclusion is obtained for the case $3\times 8=24$ events ($3\times$ 2015 NOvA conditions) that is slightly far from
the median expectations (at $\delta_{CP}=1.5\,\pi$, $3\times 5.3=15.8$ and $3\times 6.9=20.8$ are expected for IH and NH, respectively).
Including the systematic errors, scaled from the 2015 NOvA analysis, the achievement is not spoiled~\cite{lucas-mh}, with
about a decrease of 0.3-0.4 $\sigma$ in significance.

\begin{figure}[h]
\centering
\includegraphics[width=9cm]{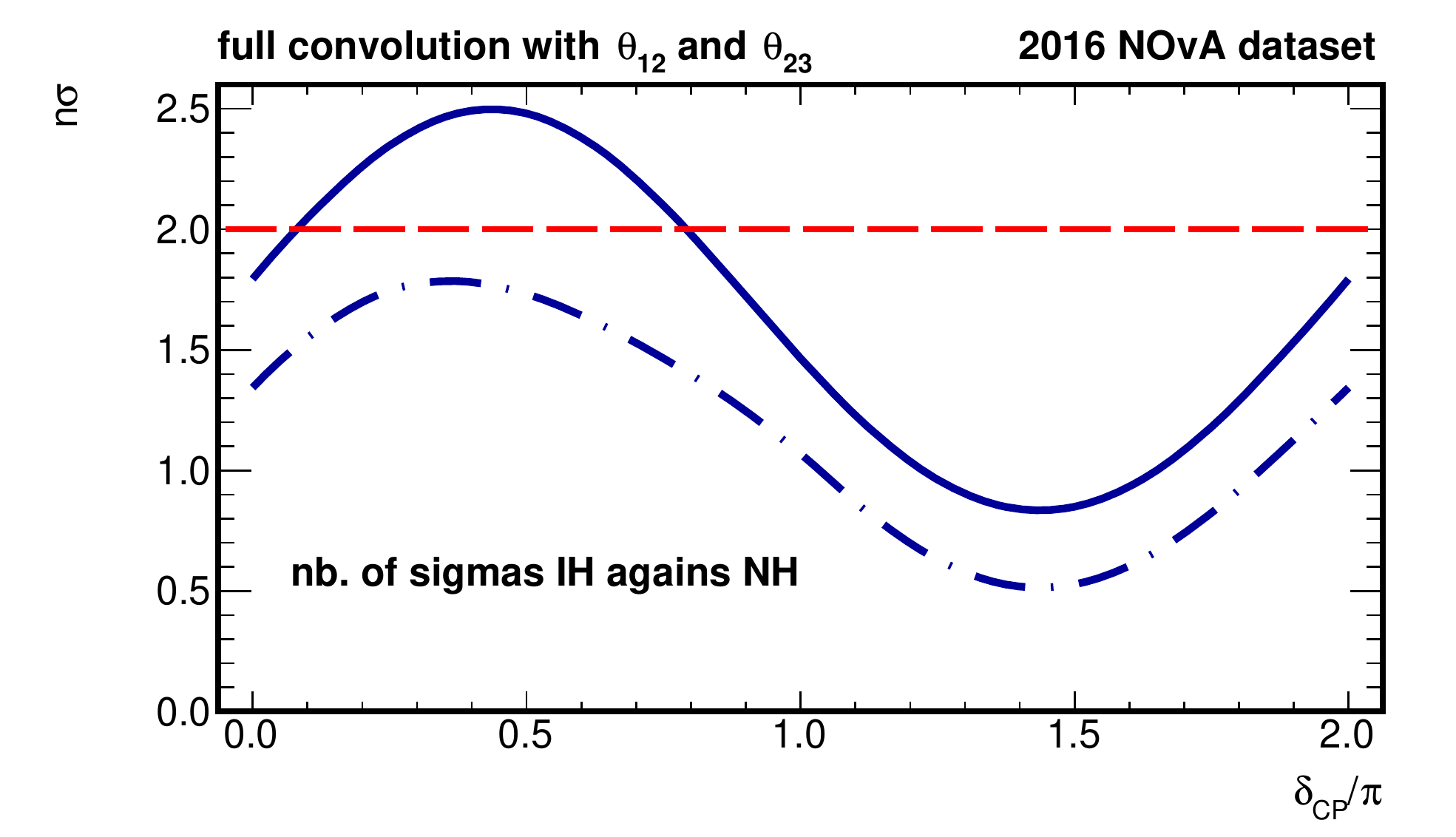}
\caption{\label{fig-9} The exclusion of the inverted hierarchy as obtained by the new statistical method
applied to the recent release of NOvA 2016 results. The uncertainties on  $\theta_{23}$ and $\theta_{13}$ are treated as nuisances,
i.e. with a full convolution of their probability densities.
IH is rejected at 95\% C.L. in the $\delta_{CP}$ interval [0.10 $\pi$, 0.77 $\pi$]. 
The standard analysis with $\Delta\chi^2_{min}$ is also shown as dashed-dot line. Taken from~\cite{lucas-mhv2}.
}
\end{figure}

NOvA collaboration just released preliminary new results~\cite{nova-prel} obtained from a total exposure of 
$6.05\times 10^{20}$ p.o.t., a factor 2.2 with regard to the 2015 results.
33 events for \numunue appearance production were found. However the background level was highly enhanced
(factor 4.5) against an increase of the efficiency of a factor 2.5. 
By scaling these number to the 2015 analysis and exposure, the 33 events of 2016 corresponds 
to about 6 events in 2015. That is  around the median expectation without an even moderate fluctuation. 
Any how, applying the new method~\cite{lucas-mhv2}, the gain in exposure from 2015 to 2016 is sufficient to obtain a first important result:
the inverted hierarchy can be excluded at 95\% C.L. in the $\delta_{CP}$ interval [0.10 $\pi$, 0.77 $\pi$] (Fig.~\ref{fig-9}).

\section{Conclusions}

The neutrino mass ordering (MO) is one of the most relevant questions to be answered in the near future. Despite the huge
investments in running experiments (NOvA and T2K) and future projects (JUNO and ORCA/PINGU at first level) the real
possibility to reach a robust result on MO is still too low, even in 10 years from now. Its strong
correlation with the $\delta_{CP}$ phase contributes to puzzling the scenario.
It is mandatory to explore new ways 
in statistical analysis, also motivated by the too simple methods applied so far, all based on the minimization of the $\chi^2$.
The new method introduced in~\cite{lucas-mh} is quite promising. In the next few years, working out first the NOvA data sample,
it would provide reliable results on the mass hierarchy, within the framework of the 3--flavour neutrinos. 
If the future NOvA results would be in line with its 2015 result where some fluctuation from the median expectations were observed, 
and assuming the three-neutrino 
oscillation paradigm be true, then the inverted hierarchy could be disproved at 95\% C.L. in the full range of $\delta_{CP}$.
Meanwhile, with the preliminary 2016 NOvA data release, the new method allows to exclude IH at 95\% C.L. in the $\delta_{CP}$ 
interval [0.10 $\pi$, 0.77 $\pi$]. 
Moreover, plans of the authors are to add T2K data and to extend their technique to the next projects, JUNO and ORCA/PINGU.

To conclude it is time to look forward and evaluate the consequences of a possible MO knowledge to the future neutrino programs:
\begin{itemize}
\item T2K would have an easier life for the first measurement of $\delta_{CP}$;
\item JUNO would become relevant mainly for checking the consistency of the $3\nu$ framework, since it will measure
MO almost in vacuum conditions, without the $\delta_{CP}$ dependence;
\item DUNE/HyperK may be forced to partially re-tune their goals.
\end{itemize}

\label {conc}

\begin{acknowledgement}
I wish to acknowledge the kind invitation to present for the first time publicly our new results on the neutrino mass ordering.
I am also delightful of the warm hospitality of the organizers of ICFNP 2016. I also acknowledge R. Brugnera and S. Dusini
for the proof-reading of the manuscript and providing relevant criticisms.
\end{acknowledgement}

%
%
%

\end{document}